\documentclass[USenglish,twocolumn]{article}	

\usepackage[utf8]{inputenc}				
\usepackage[big,online]{dgruyter}	
\usepackage{lmodern} 
\usepackage{microtype}
\usepackage[numbers,square,sort&compress]{natbib}

\usepackage{hyperref}
\usepackage{amsfonts}
\usepackage{amsmath,etoolbox}

\usepackage{amsbsy}
\usepackage{xcolor}
\usepackage{soul}
\usepackage{graphicx}
\usepackage{mathtools}

\newcommand{\ket}[1]{\lvert#1\rangle}

\newcommand{\abst}[1]{\lvert#1\rvert^2}

\newcommand{\Pc}{\mathcal{P}}
\newcommand{\cbE}{\boldsymbol{\mathbf{\cal E}}}

\newcommand{\unitvec}[1]{\hat{\mathbf{{#1}}}}
\renewcommand{\vec}[1]{\mathbf{#1}}

\usepackage{float}

\begin{document}

	\articletype{Research Article}
	\received{Month	DD, YYYY}
	\revised{Month	DD, YYYY}
  \accepted{Month	DD, YYYY}
  \journalname{De~Gruyter~Journal}
  \journalyear{YYYY}
  \journalvolume{XX}
  \journalissue{X}
  \startpage{1}
  \aop
  \DOI{10.1515/sample-YYYY-XXXX}

\title{Cooperative optical wavefront engineering with atomic arrays}
\runningtitle{Cooperative optical wavefront engineering with atomic arrays}

\author[2]{K.~E.~Ballantine}
\author[1]{J.~Ruostekoski}

\affil[1]{\protect\raggedright  Department of Physics, Lancaster University, Lancaster, LA1 4YB, United Kingdom, E-mail: j.ruostekoski@lancaster.ac.uk}
\affil[2]{\protect\raggedright  Department of Physics, Lancaster University, Lancaster, LA1 4YB, United Kingdom, E-mail: k.ballantine@lancaster.ac.uk}

\abstract{
Natural materials typically interact weakly with the magnetic component of light which greatly limits their applications. This has led to the development of artificial metamaterials and metasurfaces. However, natural atoms, where only electric dipole transitions are relevant at optical frequencies, can cooperatively respond to light to form collective excitations with strong magnetic, as well as electric, interactions, together with corresponding electric and magnetic mirror reflection properties. By combining the electric and magnetic collective degrees of freedom we show that ultrathin planar arrays of atoms can be utilized as atomic lenses to focus light to subwavelength spots at the diffraction limit, to steer light at different angles allowing for optical sorting, and as converters between different angular momentum states. The method is based on coherently superposing induced electric and magnetic dipoles to engineer a quantum nanophotonic Huygens' surface of atoms, giving full $2\pi$ phase control over the transmission, with close to zero reflection. }

\maketitle

\section{Introduction}

The quest for artificial materials with a strong magnetic, as well as electric, response at optical frequencies, has fueled the rich and rapidly expanding field of metamaterials~\cite{Zheludev12}. Thin, effectively two-dimensional (2D), layers of such metamaterials, known as metasurfaces, can impart an abrupt phase shift on transmitted or reflected light, allowing for unconventional beam shaping over subwavelength distances~\cite{Yu14,Luo18}. An important example of a metasurface is the Huygens' surface, based on Huygens' principle, that every point acts as an ideal source of forward propagating waves~\cite{Huygens,Love1901}. By engineering crossed electric and magnetic dipoles, a physical implementation of Huygens' fictitious sources can be realized, providing full transmission with arbitrary $2\pi$ phase allowing extreme control and manipulation of light~\cite{Pfeiffer13,Decker15,Yu15,Chong15,Shalaev15}.

The use of artificial metamaterials for these applications is due to the restriction that most natural materials interact weakly with magnetic fields at optical frequencies, to the extent that the magnetic response can be considered negligible. However, arrays of atoms with a dominant electric dipole transition can have \emph{collective} excitations which interact strongly with the magnetic light component~\cite{Ballantine20Huygens,Alaee20}. As the ability to use the electric dipole response of regular ultrathin 2D planar arrays of atoms to control and manipulate light has already been explored in many contexts~\cite{Jenkins2012a,Bettles2017,Perczel2017a,Plankensteiner2017,Facchinetti16,Bettles2016,Yoo2016,Asenjo-Garcia2017a,Mkhitaryan18,Orioli19,Shahmoon,Javanainen19,
Bettles20,Parmee2020}, this opens the way to combining both electric and magnetic degrees of freedom. Subwavelength arrays of atoms can operate at a single-photon quantum level~\cite{Guimond2019,Williamson2020b,Cidrim20,
Ballantine20ant,Zhang21,Cardoner21} and are increasingly experimentally achievable~\cite{Rui2020,Glicenstein2020,Endres16,Barredo18}. Indeed, recent experiments have already investigated the cooperative response of a single layer of atoms, and found characteristic spectral narrowing below the fundamental quantum limit for a single atom~\cite{Rui2020}.

Here we illustrate how an atomic Huygens' surface~\cite{Ballantine20Huygens} can be used for novel beam-shaping and optical manipulation applications. Collective coherent, uniform electric-dipole and magnetic-dipole excitations of atomic arrays have different fundamental reflection properties, corresponding to electric and magnetic mirrors, for which in the latter case the standard $\pi$ phase shift of the reflected beam is absent.
We show how superpositions of such collective excitations can be used to create a nearly reflection-less Huygens' surface to engineer the wavefront of the light, so that the atomic array acts as an ultrathin flat lens, an optical sorter, or a converter to different orbital angular momentum (OAM) states of light. The simulations of the atomic array produce diffraction-limited focusing of light with very short wavelength-scale focal lengths. The optical sorting is achieved by steering the beam's propagation in desired directions using the atoms with negligible reflection.

\section{Engineering optically active magnetism}

\subsection{Atom-light interaction}

\begin{figure}[htbp]
  \centering
   \includegraphics[width=\columnwidth]{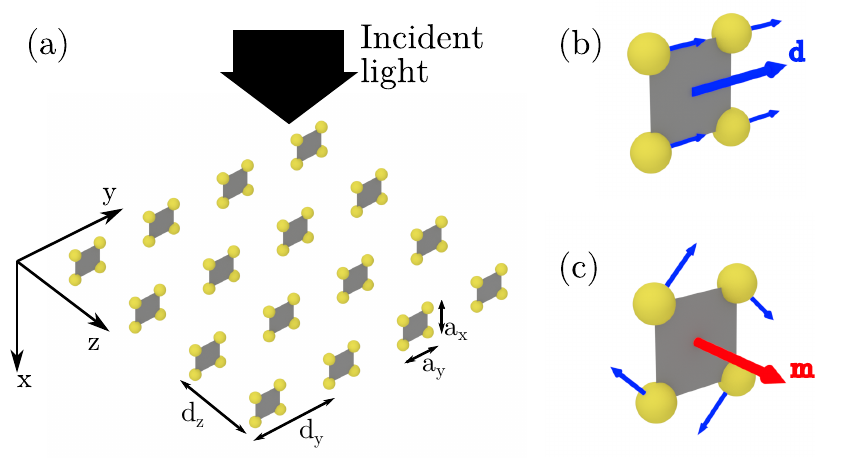}
   \vspace{-0.6cm}
  \caption{An atomic Huygens' surface with strong magnetic response at optical frequencies.
  (a) The atomic array consists of a 2D square lattice in the $yz$ plane. Each site further consists of a square unit cell of four atoms, forming an atomic bilayer. (b) Uniform polarization on each atom leads to an effective electric dipole moment $\vec{d}$ from the unit cell. (c) Azimuthal polarization leads to a net zero electric dipole moment, but a perpendicular magnetic dipole moment $\vec{m}$. 
  }
  \label{fig:fig1}
  \end{figure}

We consider a rectangular lattice in the $yz$ plane, at $x=0$, with spacing $d_y$, $d_z$. Every lattice site consists of a unit cell of four atoms displaced by $\pm(a_x/2)\unitvec{x}\pm (a_y/2) \unitvec{y}$, with a $\ket{J=0}\rightarrow\ket{J^\prime=1,m=\sigma}$ transition. Such a geometry, illustrated in Fig.~\ref{fig:fig1}, could be realized as a bilayer optical lattice~\cite{Koepsell20,Gall21}, with a double-well superlattice in the $y$ direction with two minima at $\pm(a_y/2)\unitvec{y}$ in every period $d_y$, or by optical tweezers~\cite{Barredo18}. Each atom is driven by the coherent incident laser field $\cbE={\cal E}(\vec{r}) e^{ikx}\unitvec{e}_y$ (all field components and amplitudes here and in the following refer to the slowly varying, positive frequency components with the rapid variations $\sim\exp{(i\Omega t)}$ at the laser frequency $\Omega$ filtered out), as well as by the scattered field from all other atoms. In the limit of a low-intensity coherent laser-drive the atoms behave as classical linear coupled dipoles, i.e.\ as a set of damped, driven, coupled, harmonic oscillators, as described in detail in Refs.~\cite{Lee16,Morice1995a,Ruostekoski1997a,
Sokolov2011}. The origin of these dipoles are the quantum-mechanical electronic transitions between atomic orbitals. Due to the selection rules, only resonant transitions corresponding to electric dipoles can be relevant, while other multipoles, including magnetic dipoles, are negligible. The total electric field amplitude is then the sum of the incident field amplitude and the scattered field from each atom acting as a point electric dipole~\cite{Jackson}, and from the resulting coupled equations the light field can be solved
exactly~\cite{Lee16}. For subwavelength spacing, the light-mediated long-range interactions between the atoms are strong due to recurrent multiple scattering where a photon is scattered more than once by the same atom. 
The steady-state of the system in the low light-intensity limit is described by the set of polarization amplitudes $\Pc_{\sigma}^{(j)}$, where
the dipole moment of atom $j$ is $\vec{d}_j=\mathcal{D}\sum_\sigma \Pc_\sigma^{(j)}\unitvec{e}_\sigma$, and $\mathcal{D}$ and  $\unitvec{e}_\sigma$ are the reduced dipole matrix element and the relevant unit vector, respectively~\cite{Lee16}.  An identical formalism, without drive, describes the decay of a single-photon excitation in the absence of an incident laser field, where $\Pc_\sigma^{(j)}$ then represents the probability amplitude of the excitation on level $\sigma$ of atom $j$~\cite{SVI10,Ballantine20ant}. 

The response of the array to incident light can be understood in terms of the collective excitation eigenmodes $\vec{v}_n$ of the radiatively coupled atoms, with eigenvalues $\delta_n+i\upsilon_n$, where $\delta_n$ is the collective line shift and $\upsilon_n$ is the collective linewidth, which can differ dramatically from the linewidth $\gamma$ of a single atom~\cite{Jenkins_long16}. It is useful to define the biorthogonality occupation measure of the collective eigenmode $\vec{v}_j$ as~\cite{Facchinetti16}
\begin{equation}
L_j=\frac{\abst{\vec{v}_j^T\vec{b}}}{\sum_k\abst{\vec{v}_k^T\vec{b}}},
\end{equation}
where $\vec{b}_{\sigma+3j-1}=\Pc_\sigma^{(j)}$ denote the polarization amplitudes in vector form.

To understand the collective excitations of the entire array, we first consider an isolated unit cell of four atoms. For $a\lesssim \lambda$, the scattered field from an individual unit cell can be well described by a multipole decomposition in terms of single electric and magnetic dipoles, quadrupoles, etc., at its center. Each unit cell in isolation exhibits collective eigenmodes, owing to the light-mediated coupling between the four atoms. The eigenmode shown in Fig.~\ref{fig:fig1}(b), consists of all atomic dipoles oscillating in phase and pointing in one direction leading to an effective electric dipole moment. While this collective mode replicates the electric dipole moment of individual atoms, radiative excitations with, e.g., magnetic properties, not present in individual atoms, can also be engineered by utilizing more complex collective excitation eigenmodes. 

A second eigenmode, shown in Fig.~\ref{fig:fig1}(c), consists of an arrangement of four atoms at the corners of a rectangle, representing a net zero electric dipole, but a perpendicular magnetic dipole. The orientations of quantum-mechanical atomic transitions in these four point-like discrete atoms, each of which generates an electric dipole,
approximate  a circular loop of a continuous azimuthal electric polarization density. In this collective eigenmode of Fig.~\ref{fig:fig1}(c), the electric dipoles at each point due to the electronic transitions therefore leads to an equivalent circulating current around the center producing a magnetic dipole moment, analogous to a classical continuous distribution of oscillating charges on a ring. This subwavelength current loop is close to indistinguishable in the far-field from a fundamental magnetic dipole~\cite{Ballantine20Huygens,Alaee20}. 

Strong light-mediated interactions between different unit cells then lead to collective radiative excitations of the whole lattice that synthesize effective electric and magnetic dipole arrays. In particular, the lattice in Fig.~\ref{fig:fig1}(a) has two collective excitation eigenmodes of interest. One is a uniform repetition of coherent, in-phase, electric dipole moments oriented along the $y$ direction on each unit cell, and has a collective linewidth comparable with a single atom~\cite{Facchinetti18}. The other is a similarly coherent, uniform, excitation of magnetic dipole moments oriented along the $z$ direction, which has a significantly narrower linewidth for large lattices.

\subsection{Magnetic Mirror}

Full reflection from an array of dipoles can occur when $\vec{E}_s^{(+)}=-\cbE$, where $\vec{E}_s^{(+)}$ ($\vec{E}_s^{(-)}$) is the scattered field in the forward (backward) direction, leading to destructive interference in the transmitted light and a standing wave in the backward direction. As seen in Fig.~\ref{fig:fig1}(b), the electric dipole collective excitation eigenmode has polarization amplitude that is symmetric around $x=0$, and so $\vec{E}_s^{(-)}=\vec{E}_s^{(+)}$. 
Defining the complex reflection and transmission amplitudes,
\begin{equation}
r=\frac{\unitvec{e}_y\cdot\vec{E}_s^{(-)}}{\unitvec{e}_y\cdot\cbE},\quad t=\frac{\unitvec{e}_y\cdot\left[\cbE + \vec{E}_s^{(+)}\right]}{\unitvec{e}_y\cdot\cbE},
\end{equation}
respectively, gives $r=-1$ when $t=0$. While here we consider a collective mode of effective dipoles formed on each unit cell,  this result is well-known for uniform arrays of coherently oscillating electric dipoles~\cite{ Tretyakov,Abajo07,Moitra2015,CAIT}, including equivalent single layers of atoms~\cite{Bettles2016,Facchinetti16,Shahmoon}. The $r=-1$ condition leads to a node of the standing wave at the array, equivalent to reflection from a perfect electrical conductor. Indeed, reflection from the electric-dipole collective excitation of a single-layer atomic array with subwavelength spacing has now been observed in experiments~\cite{Rui2020}. 
An alternative condition, with $r=1$, occurs for reflection from magnetic dipole excitations in metal~\cite{Sievenpiper99}, dielectric~\cite{Ginn12,Liu14,Lin16}, and atomic structures~\cite{Alaee20}. For the magnetic dipole excitation shown in Fig.~\ref{fig:fig1}(c), the $y$ component of polarization is anti-symmetric in $x$. For the uniform, coherent, excitation of a large lattice with all dipoles oscillating in phase, the scattered field in the forward and backward directions depends only on this in-plane component, and $\vec{E}_s^{(-)}=-\vec{E}_s^{(+)}$. Hence, when the magnetic dipole lattice displays full reflection it is with $r=1$, and an antinode at the plane of the mirror. This distinction means that an emitter placed in the near-field will interfere constructively with its image, rather than destructively as for an electric mirror~\cite{Liu14}.

\begin{figure}[htbp]
  \centering
   \includegraphics[width=\columnwidth]{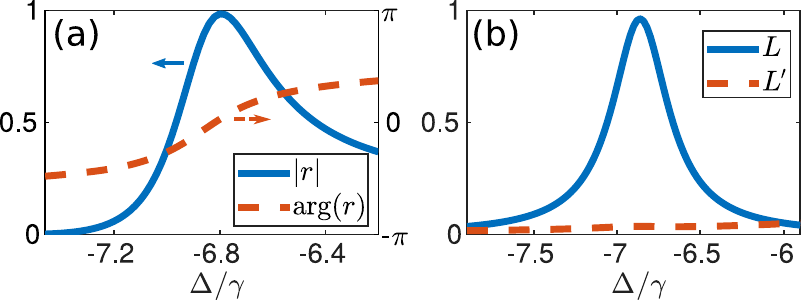}
   \vspace{-0.6cm}
  \caption{ Collective resonance of a magnetic mirror.
  (a) Magnitude (left axis) and phase (right axis) of reflection coefficient $r$ showing $r\approx 1$ close to magnetic resonance. (b) Occupation $L$ of magnetic dipole mode and $L^\prime$ of all other modes, normalized such that $\sum_j L_j=1$ at the center of the resonance. Gaussian beam with ${\cal E}(0)=1$, beam waist radius $w_0=6.4\lambda$, incident on $25\times 25 \times 4$ lattice with $d_y=d_z=0.7\lambda$,  $a_x=a_y=0.15 \lambda$.
  }
  \label{fig:mm2}
  \end{figure}

Figure~\ref{fig:mm2}(a) shows the magnitude and phase of the reflection amplitude for an incident Gaussian beam, as the detuning $\Delta=\Delta_\sigma^{(j)}=\Omega-\omega^{(j)}_\sigma$ of the laser frequency from the identical atomic resonance $\omega_\sigma^{(j)}$ of each level $\sigma$ on atom $j$, is tuned through the resonance of the collective magnetic excitation. While the magnetic dipole mode is very subradiant for large lattices, it can be well excited by an incident plane wave or Gaussian because the $\exp{(ikx)}$ rapid phase variance in the $x$ direction has an overlap with the antisymmetric polarization amplitudes of the mode. We find the numerical value $r\approx 0.99\exp{(0.1i)}$ on resonance, close to the expected value $r= 1$. This full reflection is a result of the collective nature of the excitation, and depends on the uniform, in-phase, coherent nature of the magnetic dipole  oscillations, as well as the symmetry of the individual dipole radiation. The occupation $L$ of the collective excitation eigenmode corresponding most closely to this ideal uniform magnetic dipole excitation is shown in Fig.~\ref{fig:mm2}(b), along with the occupation of all other modes. Although the magnetic mode is dominant, it is the small contribution from other modes, with different symmetry, which leads to a minor deviation from $r=1$.  
  
\begin{figure}[htbp]
  \centering
   \includegraphics[width=\columnwidth]{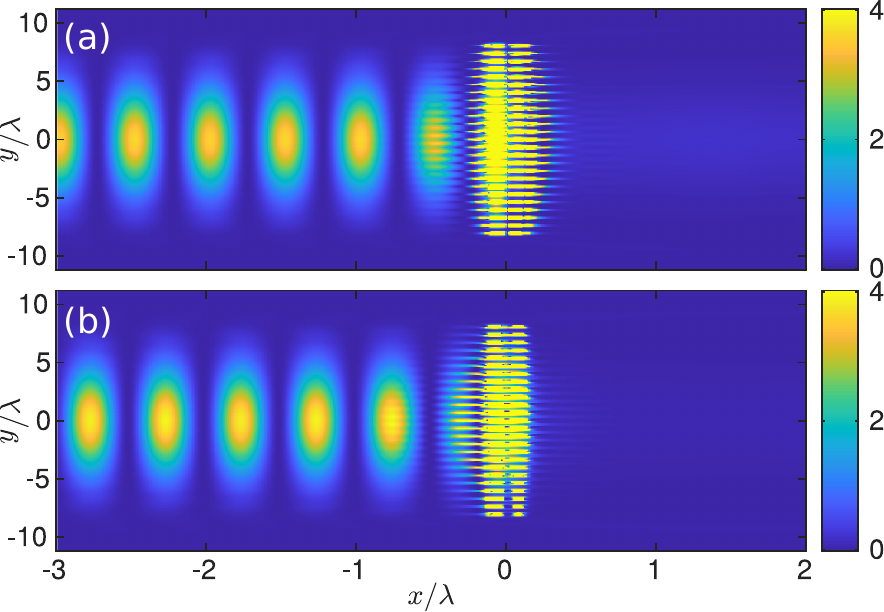}
   \vspace{-0.6cm}
  \caption{Magnetic and electric mirror formed by a bilayer of atoms. Spatial variation of $|\vec{E}|^2/|{\cal E}(0)|^2$ in $z=0$ plane,
resulting from interference between incoming Gaussian beam and reflected light from (a) magnetic mirror, when the incident light is resonant with the collective magnetic dipole mode, showing antinodes at integer and half integer distances from the lattice corresponding to $r\approx 1$, and (b) on resonance with the electric dipole mode with $r\approx -1$. Parameters as in Fig.~\ref{fig:mm2}.
  }
  \label{fig:mm1}
  \end{figure}
  
In Fig.~\ref{fig:mm1} the spatial dependence of the intensity of the total (incident plus scattered) light is shown at the frequency of the collective magnetic mode, and at the collective electric dipole mode resonance. As well as almost complete reflection, a clear $\lambda/4$ shift in the positions of the peaks of the resulting standing wave is evident.

\section{Huygens' surface}

We superpose the collective electric-dipole and magnetic-dipole excitations to form a nearly reflection-less Huygens' surface that controls the phase of the transmitted light to engineer its wavefront. A Huygens' surface is a physical implementation of
Huygens' principle~\cite{Huygens}, which states that every point in a propagating wave acts as a source of further forward-propagating waves. While electric and magnetic dipoles both scatter light forwards and backwards with equal amplitude, a crossed electric and magnetic dipole can lead to destructive interference in the backward direction and constructive interference in the forward direction, providing a physical realization of Huygens' fictitious sources~\cite{Love1901,Schelkunoff36}. 

\begin{figure}[htbp]
  \centering
   \includegraphics[width=\columnwidth]{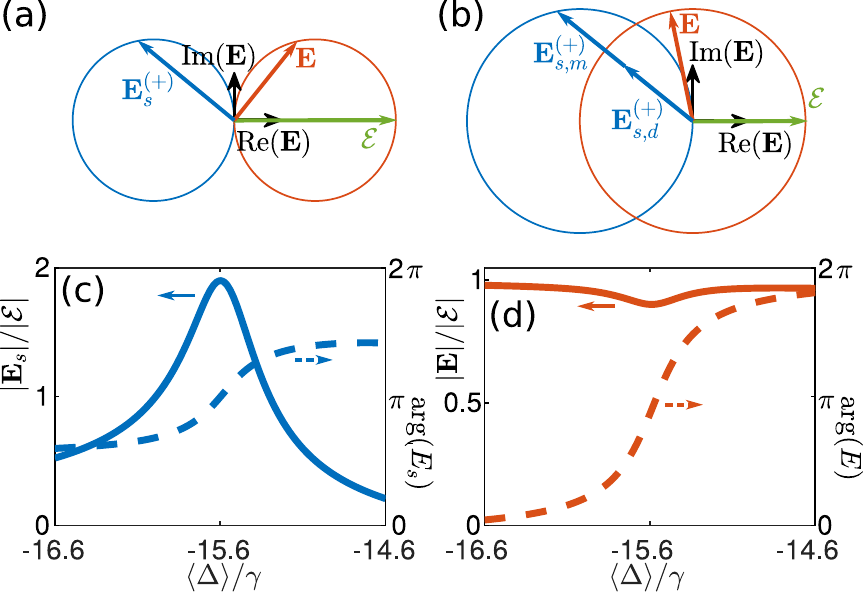}
   \vspace{-0.6cm}
  \caption{
  Principle of Huygens' surface of atoms. Possible complex values for the forward scattered field $\vec{E}_s^{(+)}$ and the total transmitted field $\vec{E}=\cbE + \vec{E}_s^{(+)}$ due to an incident field $\cbE$ for (a) a uniform electric dipole excitation and (b) crossed electric and magnetic dipoles at each site. Circles show range of possible values traced out as the detuning is varied. (c) Numerical calculation of  scattered field from plane wave (${\cal E}=1$) incident on an atomic Huygens' surface as the detuning $\langle\Delta\rangle$ of the laser from the average single-atom resonance is varied. The phase of the scattered light (dashed line, right axis) ranges from $\pi/2$ to $3\pi/2$ as a function of the laser frequency, while the magnitude (solid line, left axis) can be almost double the input field. (d) Magnitude (left axis) and phase (right axis) of total forward-propagating field showing high transmission over the full $2\pi$ range. (c,d) Are for $20\times 20\times 4$ lattice with $d_y=d_z=0.8\lambda$, $a_x=0.12\lambda$, $a_y=0.11\lambda$.   
  }
  \label{fig:hexp}
  \end{figure}

The principle of how simultaneous excitation of both modes can lead to full transmission, with arbitrary phase, is illustrated in Fig.~\ref{fig:hexp}. In Fig.~\ref{fig:hexp}(a), the individual response of either of the collective excitation eigenmodes (electric or magnetic) to an incident field $\cbE$ is shown, where the arrows display the forward-scattered field $\vec{E}_s^{(+)}$ and the total transmitted field $\vec{E}=\cbE+ \vec{E}_s^{(+)}$ in the complex plane for some particular detuning from resonance of the incident laser frequency, and the circles illustrate the range of possible values these fields can take as the detuning varies. As resonance is approached the scattered field grows originally at a phase $\pi/2$ to the incident field, reaching a maximum amplitude when $\vec{E}_s^{(+)}=-\cbE$, and then falling off as the phase approaches $3\pi/2$, tracing out the blue circle. The total transmitted field $\vec{E}$ meanwhile, starts equal to $\cbE$ far from resonance, then decreases to zero where the phase shifts from $\pi/2$ to $-\pi/2$, before returning to $\cbE$, tracing out the orange circle. The result is $-\pi/2 < \mathrm{arg}(t)<\pi/2$, and $|t|\leq 1$, with $|t|=0$ on resonance. 

For the simultaneous excitation of the uniform collective modes corresponding to crossed magnetic and electric dipoles, illustrated in Fig.~\ref{fig:hexp}(b), we write $\vec{E}_s^{(\pm)}=\vec{E}_{s,d}^{(\pm)}+\vec{E}_{s,m}^{(\pm)}$ where $\vec{E}_{s,d}^{(\pm)}$ ($\vec{E}_{s,m}^{(\pm)}$) is the contribution from the electric (magnetic) dipoles. Again, away from resonance, $\vec{E}_s^{(+)}=0$ and $t=1$. As the resonance is approached, however, the two contributions lead to double the amplitude, and on resonance $\vec{E}_{s,m}^{(+)}=\vec{E}_{s,d}^{(+)}=-\cbE$, giving $t=-1$, after which the scattered field decreases again to zero. In between the phase of $\vec{E}$ varies over the full $2\pi$ phase range while the symmetry condition $\vec{E}_{s,d}^{(-)}=-\vec{E}_{s,m}^{(-)}$ ensures $r=0$ as expected~\cite{Love1901}. The destructive interference of the reflected fields, which is a pre-condition for a Huygens' surface leading to full transmission,  is known as the Kerker effect~\cite{Kerker83,Liu18}.

The numerically calculated scattered light from the atomic array in Fig.~\ref{fig:hexp}(c) shows that indeed the scattered field approaches twice the magnitude of the incident field. While the phase of the scattered field is constrained between $\pi/2$ and $3\pi/2$, as is the case for a single uniform excitation, this increased magnitude means that the total field in Fig.~\ref{fig:hexp}(d) covers a full $2\pi$ phase range while the transmission $|t|=|\vec{E}|/|\mathcal{E}|$ remains close to one. We find from a multipole decomposition that at the center of the resonance the normalized electric dipole moment from a single unit cell is $0.55$, while the magnetic dipole moment is $0.44$, with the remaining contribution from quadrupole or higher moments.  

To simultaneously excite the electric and magnetic collective excitation eigenmodes, we take the level shifts $\Delta_\sigma^{(j)}$ to vary independently between atomic levels within a unit cell, while keeping them identical between each unit cell. The relative level shifts, along with the unit cell size and lattice constants, are numerically optimized to maximize transmission, and here in the studied examples vary between $\pm 4\gamma$.
The level shifts could be achieved by ac Stark shifts~\cite{gerbier_pra_2006} of standing waves, with similar periodicity as the unit cells of the lattice, such that the pattern is repeated on each cell, while for a slow variation across the array also microwave or magnetic fields can be suitable. 

The result of the numerical optimization we employ here can be understood as inducing a coupling between the superradiant collective mode of uniform electric dipoles on each unit cell, with $\upsilon=1.3\gamma$, and the subradiant collective mode of uniform magnetic dipoles, with $\upsilon=0.1\gamma$. For a single atom, although the incident light directly only drives the atomic dipoles along the $y$ direction, relative level shifts of the $m=\pm1$ states lead to coupling between $x$ and $y$ polarization amplitudes. Then, over the entire lattice, repeated patterns of varying level shifts can be used to engineer an effective coupling between collective modes of the array, allowing modes to be occupied even if they are not directly driven~\cite{Facchinetti16,Facchinetti18,Ballantine20Huygens,Ballantine20Toroidal}.
The maximum magnitude of scattered field in Fig.~\ref{fig:hexp}(c) occurs when the incident light is on resonance with the magnetic dipole mode, which as a result is highly occupied with $L=0.88$. The resulting scattered field from the magnetic dipoles again approximately cancels the incident field, with $\vec{E}_{s,m}^{(+)}\sim -\cbE$, as in the magnetic mirror case shown in Fig.~\ref{fig:mm1}. However, the coupling also leads to a small occupation, $L=0.08$, of the off-resonance electric dipole mode. The smaller occupation of this mode is compensated by the larger linewidth, leading to a similar magnitude of scattered field with $\vec{E}_{s,d}^{(+)}\sim \vec{E}_{s,m}^{(+)}\sim -\cbE$, and $\vec{E}_{s,d}^{(-)}\sim -\vec{E}_{s,m}^{(-)}$. The result is close to full transmission with a $\pi$ phase shift. As the occupation of the electric dipole mode is a direct result of coupling from the magnetic mode, the occupation of both modes falls away on either side of the magnetic resonance, and the total scattered field decreases to zero.

\section{Beam focusing}

\begin{figure}[htbp]
  \centering
   \includegraphics[width=\columnwidth]{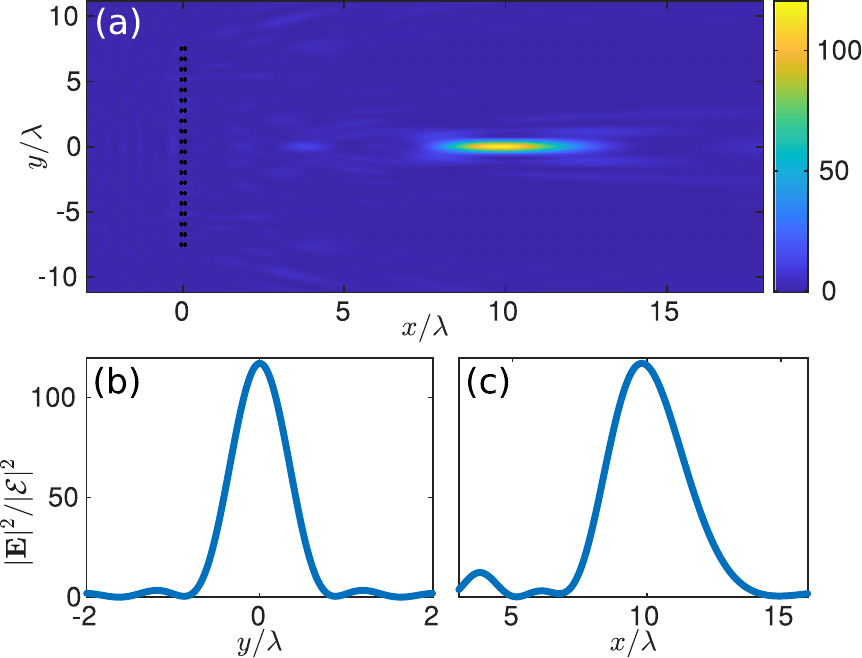}
   \vspace{-0.6cm}
  \caption{
  Beam focusing by ultrathin atomic lens at a focal length $f=10\lambda$. (a) $|\vec{E}|^2/|{\cal E}|^2$ from plane wave ${\cal E}=1$ incident on lattice at $x=0$. (b) Focal spot along $y$ axis at $x=10\lambda$, $z=0$ and (c) along $x$ axis at $y,z=0$. Focal spot has FWHM $0.77\lambda$ in plane and $4.0\lambda$ along $x$ axis. Lattice parameters as in Fig.~\ref{fig:hexp}. Atomic positions are illustrated by black dots.
  }
  \label{fig:focus}
  \end{figure}
  
We show how an ultrathin flat atomic array,  acting as a Huygens' surface, can achieve tight diffraction-limited focusing in a propagation distance of a few wavelengths. The transmitted phase profile is chosen to be equivalent to a lens, namely~\cite{Aieta12,Khorasaninejad16}
\begin{equation}
\phi = \left(\frac{2\pi}{\lambda}\right)\left(f-\sqrt{f^2+\rho^2}\right),
\end{equation}
where $f$ is the focal length of the lens and $\rho=\sqrt{y^2+z^2}$, but because of the abrupt phase change at the array, this is achieved in a vastly smaller propagation distance than a solid lens. Here, the phase profile is implemented by adding a slowly-varying spatially dependent shift to all levels, in the range $\pm\gamma$, such that the locally transmitted phase varies as predicted by Fig.~\ref{fig:hexp}(d). In free space the spot size is limited by the Abbe-Rayleigh diffraction limit,
$
\lambda/\left(2\sin{\theta}\right),
$
where $\theta$ is the maximum angle of incoming rays, such that $\tan{\theta} = L/(2f)$ for an array of  side length $L$. An example is shown in Fig.~\ref{fig:focus}, for $f=10\lambda$. The resulting spot size has a subwavelength full-width-at-half-maximum (FWHM) of $0.78\lambda$, approximately equal to the diffraction limit in this case. Focusing close to the diffraction limit can be achieved for a range of focal lengths, down to $f\approx\lambda$. More complex phase variations could, in principle, be used to correct aberrations and further limit the focal spot size as has been achieved in plasmonic and dielectric metalenses~\cite{Liu20}.

\section{Beam steering}

\begin{figure}[htbp]
  \centering
   \includegraphics[width=\columnwidth]{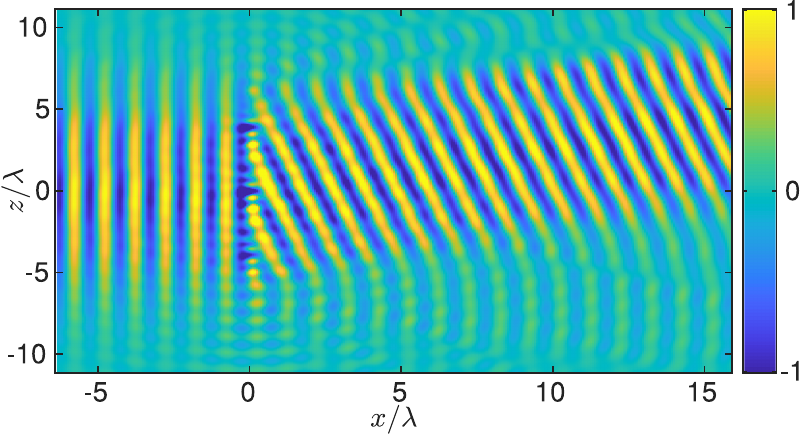}
   \vspace{-0.6cm}
  \caption{
  Beam steering by $15^\circ$ by using an atomic Huygens' surface. Real part of electric field $\mathrm{Re}(E_y)/|{\cal E}(0)|$ for incident Gaussian beam being steered by atomic lattice at $x=0$. Incident beam has ${\cal E}(0)=1$ and beam waist radius $w_0=6.4\lambda$.  $20\times 20 \times 4$ lattice with $d_y=0.82\lambda$, $d_z=0.65\lambda$, $a_x=0.11 \lambda$, $a_y=0.1\lambda$.
  }
  \label{fig:steering}
  \end{figure}
  
  \begin{figure*}[t!]
  \centering
   \includegraphics[width=\textwidth]{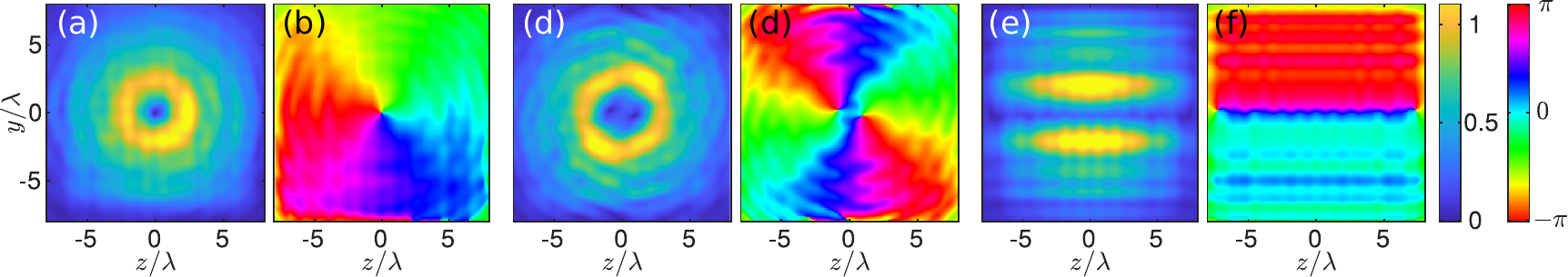}
   \vspace{-0.6cm}
  \caption{Converting an optical vortex-free beam to beams with orbital angular momentum (OAM).
 $|\vec{E}|/|{\cal E}(0)|$ (left of each pair) and $\mathrm{arg}(E_y)$ (right) for several OAM vortex beams. (a,b) $l=1$, (c,d) $l=-2$, (e,f) superposition of $l=1$ and $l=-1$, in $yz$ plane at $x=6\lambda$. Gaussian beam with ${\cal E}(0)=1$ and beam waist radius $w_0=6.4\lambda$ is incident on $20\times 20 \times 4$ lattice with parameters as in Fig.~\ref{fig:hexp}.
  }
  \label{fig:oam}
  \end{figure*}

We use an atomic array to produce beam steering, a resource in controlling the flow of light, by redirecting an incoming beam in different directions. This is achieved by adding an additional level shift gradient, which varies linearly across the lattice, such that the phase of the transmitted light also varies linearly~\cite{Pfeiffer13,Ni12,Liu17bs}.
To steer the beam by an angle $\theta$ away from the normal requires the phase profile $\phi = \alpha z$ where 
$\alpha = 2\pi\sin{\theta}/\lambda$.
For discrete dipoles with spacing $d_z$ in the $z$ direction, this phase profile is unique modulo $2\pi$ if $|\sin{\theta}| < (\lambda/d_z)-1$.
An example is shown in Fig.~\ref{fig:steering} where an incident Gaussian beam is transmitted at an angle of $15^\circ$ to the $x$ axis. Beam steering from atoms could be used to sort light for different subsequent stages, e.g., as part of a modular quantum information architecture. In contrast to fabricated plasmonic or dielectric systems, the detuning gradient can in principle be varied in-situ, changing the steering angle and allowing for the dynamic sorting and redirection of different light beams. Ultrathin atomic lattices thus offer a major advantage for beam steering, combined with focusing and other beam shaping, allowing multiple independent steps with separation on the order of wavelengths, all without losses or fabrication inconsistencies.

\section{Orbital angular momentum}

Metasurfaces can be used to generate vortex beams with OAM~\cite{Mehmood16,Zhang18}. The angular momentum of paraxial beams of light can be separated into spin angular momentum, depending on its polarization, and OAM, depending on the spatial variation of the field. Figure~\ref{fig:oam}(a-d) shows the creation of vortex beams with a phase winding $\exp{(il\phi)}$, achieved by an additional angle-dependent level shift, for integer $l=1,-2$, corresponding to a quantized OAM $l\hbar$ per photon. The incoming light, with no angular momentum, has been imparted with quantized OAM, with a corresponding torque on the atomic array.  The integer quantization of OAM provides a larger alphabet for quantum logic than typical two-state bases, a useful resource for quantum computation~\cite{Mair01}. Particularly important for quantum applications is the ability to form coherent superpositions of different OAM values. This can be achieved by matching the phase to that of the desired transmitted beam, with an example shown in Fig.~\ref{fig:oam}(e,f). While in all cases the incoming light has an intensity maximum at the center, destructive interference leads to a vortex beam with zero intensity at the beam center, and along lines of phase discontinuities. Despite the change in intensity along the beam axis, the total integrated transmission remains high, with e.g.~$|t|^2\approx 0.97$ for the $l=1$ case.

\section{Conclusion}

Strong light-mediated interactions can be used to synthesize optical magnetism in arrays of atoms. The flexibility this provides in designing collective radiative excitations opens new avenues for the control and manipulation of light, with the potential for rapid technological as well as fundamental progress. These collective excitations can be engineered by varying the atomic level shifts, leading to very different responses to incident light. Here we have shown how an atomic lattice can be used for such versatile wavefront engineering as diffraction-limited focusing, optical sorting using beam-steering, and OAM mode conversion. Atomic arrays offer advantages over plasmonic or dielectric platforms, including the absence of absorptive loss and fabrication inhomogeneities,  as well as the great flexibility to operate at the quantum limit, at the same time when the quest for developing nanophotonic quantum technologies for metasurfaces is becoming one of the key challenges in the research field~\cite{Solntsev2020}. The ultrathin nature of the arrays described here and the adaptability in wavefront engineering allows for several stages to be combined for advanced wavefront shaping and modular optical processing.  

 \begin{funding}
 We acknowledge financial support from the UK EPSRC (Grant Nos. EP/S002952/1, EP/P026133/1).
\end{funding}

\end{document}